\documentclass{WileyMSP-template}
\usepackage{amsmath}
\usepackage{indentfirst}
\usepackage{subfigure}
\usepackage{color}
\begin{document}

\pagestyle{fancy}
%\rhead{\includegraphics[width=2.5cm]{vch-logo.png}}

\title{Machine-learning-enabled vectorial opto-magnetization orientation}

\maketitle

% Author: Please give full first and last names for authors and include * after the name of all corresponding authors

\author{Weichao Yan},
\author{Zhongquan Nie*},
\author{Xunwen Zeng},
\author{Guohong Dai},
\author{Mengqiang Cai},
\author{Yun Shen} and 
\author{Xiaohua Deng}

% Dedication

%\dedication{Optional dedication here. If no dedication is required, please leave blank}

% Affiliations: Please provide adacemic titles (Prof. or Dr.) for all authors where applicable, and include an institutional email address for all corresponding authors
\begin{affiliations}
W. Yan, M. Cai, X. Deng\\
Institute of Space Science and Technology, Nanchang University, Nanchang 330031, China\\
E-mail: yanweichao@ncu.edu.cn

Z. Nie\\
Key Lab of Advanced Transducers and Intelligent Control System, Ministry of Education
and Shanxi Province, College of Physics and Optoelectronics, Taiyuan University of Technology,
Taiyuan 030024, China\\
E-mail: niezhongquan1018@163.com

G. Dai, Y. Shen\\
Department of Physics, School of Science, Nanchang University, Nanchang 330031, China\\

X. Zeng\\
Information Engineering School, Nanchang University, Nanchang 330031, China\\

\end{affiliations}

% Keywords: Please provide a minimum of three and a maximum of seven keywords, separated by commas


% Abstract should be written in the present tense and impersonal style (i.e., avoid we), and be at most 200 words long
\begin{abstract}

Manipulation of light-induced magnetization has become a fundamentally hot topic with a potentially high impact for atom trapping, confocal and magnetic resonance microscopy, and data storage. The control of the magnetization orientation mainly relies on the direct methods composed of amplitude, phase and polarization modulations of the incident light under the tight focusing condition, leaving the achievement of arbitrary desirable three-dimensional (3D) magnetization orientation complicated, inflexible and inefficient.  Here, we propose a facile approach called machine learning inverse design to achieve expected vectorial opto-magnetization orientation. This pathway is time-efficient and accurate to produce the demanded incident beam for arbitrary prescribed 3D magnetization orientation. It is highlighted that the machine learning method is not only  applied for magnetization orientations, but also widely used in the control of magnetization structures.

\end{abstract}

% Text: Please use section headings and subheadings as specified below. For communications, all section headings apart from Experimental Section should be removed
% Please make the first reference to a display item bold: \textbf{Figure 1}
% Do not abbreviate Figure, Equation, etc.; display items are always singular, i.e., Figure 1 and 2.
% Equations are always singular, i.e., Equation 1 and 2, and should be inserted using the {equation} environment, not as graphics
% Please do not use footnotes in the text, additional information can be added to the Reference list.

\section{Introduction}
Discovery of the giant magnetoresistance\cite{M1988Giant} greatly facilitated the longitudinal magnetization, enabling to accelerate the development of high capacity magnetic storage devices. In 1960s, the inverse Faraday effect (IFE)\cite{PhysRevLett.15.190} was firstly proposed to describe an optically-induced magnetization in a nonabsorbing material. Apart from the pioneering work, a new physical phenomenon called all-optical helicity-dependent switching (AO-HDS) was experimentally discovered\cite{PhysRevLett.99.047601}. Later, the behavior of AO-HDS is found that this phenomenon is not a trivial one and related to paramagnetic materials\cite{PhysRevLett.15.190} or nonmagnetic metals and antiferromagnetic metals\cite{berritta2016ab} or specific laser fluence in ferrimagnetic materials\cite{wang2018alloptical,khorsand2012role,mangin2014engineered}, magnetic thin films to multilayers and even granular films\cite{lambert2014all}. Owing to the high density, high energy efficiency and erasable feature, all-optical magnetic recording (AOMR),
has emerged as a key component for the next generation storage technology of an ultra-high areal density\cite{mangin2014engineered,lambert2014all}.
To improve the storage density, the circularly polarized beam was focused into a half-wavelength region under a high numerical aperture (NA) lens\cite{helseth2008strongly, Zhang2008High, Zhang2009Theoretical}. For the purpose of  potential practical application of the longitudinal magnetization, the magnetization needle\cite{wang2014ultralong}, magnetization chain\cite{nie2015achievement} and magnetization spot arrays\cite{nie2017three,hao2017three} were also produced by the control of the phase, amplitude, and polarization distributions of the incident beam.

~~The above achieved magnetization structures are confined to the longitudinal magnetization components, which limit the capacity of magnetization storage. Recently, a uniformly oriented in-plane magnetization was produced through focusing two counter-propagating vector beams\cite{wang2017generation}. Furthermore, arbitrary 3D orientations were achieved by diverse methods, such as the reversing electric dipoles\cite{wang2018all, Luo2019Three}, raytracing models\cite{Yan2018Arbitrarily, Yan2021Dynamic} and the axially symmetrical destruction\cite{yan2018creation}. Although the inverse designing methods can be implemented by the reversing electric dipoles and the axially symmetrical destruction, the achieved magnetization orientation is a little not coincident with these means only in the tight focusing system of an ideal $4\pi$ apparatus. One significant key is needed to find the corresponding relation between the magnetization orientation and spin orientation in realistic tight focusing condition. And the method of the raytracing models is not restricted to the ideal tight focusing system, the corresponding relation is still needed to be revealed. Apart from this factor, it is required to repeatedly search the corresponding realtions for realizing numerous different magnetization orientations. On the whole, the reported methods to obtain 3D magnetization orientations are lack of considerable flexibilities and high efficiency.
To achieve the prescribed 3D magnetization orientation, what the required incident beam is like.
%For the prescribed 3D magnetization orientation, what the required incident beam is like. 
This is an inverse question and a challenging problem, which is rarely considered in previous magnetization shapings. Nowadays, machine learning has become a powerful tool in handling various problems, for example the space weather forecast\cite{tang2020comparison} and image completion\cite{chen2019improvement}. Specially, machine learning inverse design has been demonstrated to be an accurate and time-efficient approach to solving complicated inverse problems in information storage with high density\cite{wiecha2019pushing},  metasurface imager and recognizer\cite{li2019intelligent}, and 3D vectorial holography\cite{ren2020three}. Here, we will demonstrate the generation of 3D vectorial magnetization orientation through exploiting machine learning inverse design based on multilayer perceptron artificial neural networks (MANN).

\section{Theoretical illustration for achieving 3D orientated magnetization}
\begin{figure}[htbp]
\centering
\includegraphics[width=0.6\linewidth]{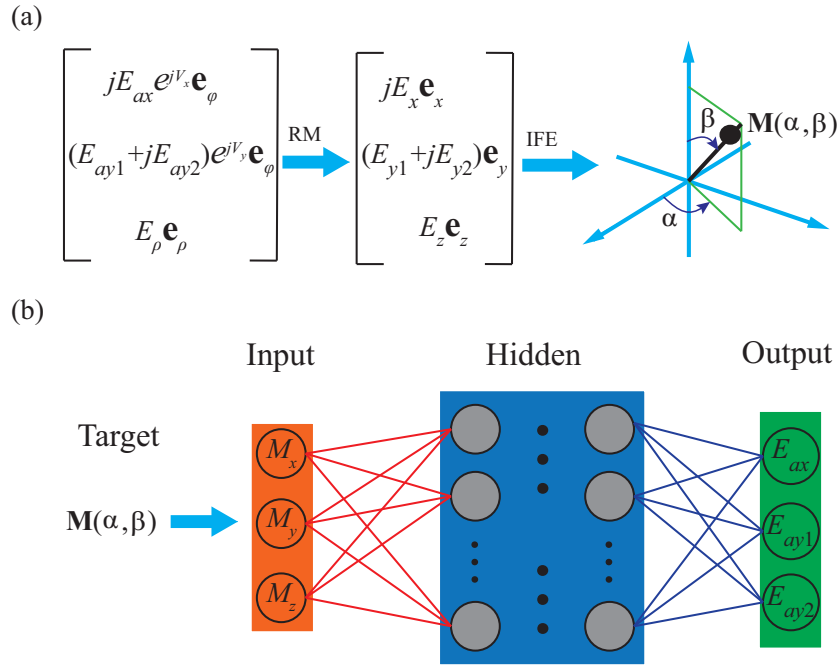}
\caption{Principle of vectorial magnetization orientation based on the machine learning. (a) Theoretical approach to achieving 3D orientated magnetization. (b) Schematic illustration of arbitrarily orientated magnetization using machine learning inverse design. RM indicates raytracing model. Input, Hidden and Output denote the input, hidden and output layers in the neural network.}
\label{fig1}
\end{figure}

The principle of vectorial magnetization orientation is described in Fig.~\ref{fig1}. 
Figure~\ref{fig1}(a) illustrates a feasible strategy for the generation of the magnetization spot with 3D orientation by the help of raytracing model\cite{Yan2018Arbitrarily, Yan2021Dynamic}. The incident beam consists of four kinds of beams, as written by the first column vector. The first kind of beam is an azimuthally polarized beam imposed with both $\pi/2$ phase delay and $\pi$-phase-step filter\cite{Ren:14} along the $x$-axis, as given by $jE_{ax}e^{jV_x}\mathbf{e}_\varphi$ in the first row of the first column vector. Here, $j$ denotes the $\pi/2$ phase delay, $e^{jV_x}$ indicates the $\pi$-phase-step filter along the $x$-axis, $E_{ax}$ represents the amplitude weight and $\mathbf{e}_\varphi$ is the azimuthal base vector. The focal field in the focus for this kind of beam can be written by $jE_x\mathbf{e}_x$ in the first row of the second column vector, where $E_x$ implies the amplitude factor and $\mathbf{e}_x$ is the Cartesian coordinate base vector along the $x$-axis. The second and third kinds of beams are azimuthally polarized beam with $\pi$-phase-step filter along the $y$-axis, and  azimuthally polarized beam imposed with both $\pi/2$ phase delay and $\pi$-phase-step filter along the $y$-axis, which are expressed as $E_{ay1}e^{jV_y}\mathbf{e}_{\varphi}$ and $jE_{ay2}e^{jV_y}\mathbf{e}_{\varphi}$ in the second row of the first column vector. Here, $E_{ay1}$ and $E_{ay2}$ are amplitude weights. Besides, $e^{jV_y}$ represents the $\pi$-phase-step filter along the $y$-axis. The focal fields in the focus for these two kinds of beams are represented by $E_{y1}\mathbf{e}_y$ and $jE_{y2}\mathbf{e}_y$ in the second row of the second column vector, respectively. Here, $E_{y1}$ and $E_{y2}$ are the corresponding amplitude factors, and $\mathbf{e}_y$ is the Cartesian coordinate base vector along the $y$-axis. The last kind of beam is a pure radially polarized beam, as written as $E_{\rho}\mathbf{e}_\rho$ in the third row of the first column vector. $E_{\rho}$ and $\mathbf{e}_\rho$ are the amplitude weight and the radial base vector, respectively. The focal field in the focus for radially polarized beam is a longitudinally polarized field written by $E_z\mathbf{e}_z$. Here, $E_z$ and $\mathbf{e}_z$ are amplitude factor and Cartesian coordinate base vector along the $z$-axis, respectively.

~~According to the general definition of the IFE proportional to the vector product between two complex conjugate fields, the combination of the focal fields such as $jE_x\mathbf{e}_x$ and $E_{y1}\mathbf{e}_y$ can induce magnetization component along the $z$-axis ($M_z$). Similarly, the combination of other focal fields can induce magnetization components along the $x$-axis and $y$-axis ($M_x$ and $M_y$). Via adjusting the weights of the mentioned four kinds of beams, the ratios between the magnetization components ($M_x$, $M_y$ and $M_z$) can be controlled, thereby achieving a magnetization spot with 3D dynamic orientation. It is pointed out that the four weights meet $E^2_{ax}+E^2_{ay1}+E^2_{ay2}+E^2_{\rho}=1$ and only three weights are completely independent. The inverse designing schematic to achieve a magnetization spot with the prescribed 3D orientation based on machine learning is depicted in Fig.~\ref{fig1}(b). The target is a vectorial magnetization written by $\mathbf{M}(\alpha, \beta)$, which is decomposed into three Cartesian orthogonal components ($M_x$, $M_y$, $M_z$). These three orthogonal components are used as the input layer. The output layer is the incident beam composed of four sorts of beams represented by independent amplitude parameters ($E_{ax}$, $E_{ay1}$ and $E_{ay2}$), where $E_\rho$ is set as 1.  The key point is to train the MANN to statistically learn the relationship between a given 3D vectorial magnetization orientation (Input) on the nonabsorbing magnetic-optic (MO) film and the 2D vector field distribution in the pupil plane (Output). 
%This relationship is built by a lot of hidden layers and each hidden layer is composed of thousands of neurons.

~~In the light of the above schematic to produce vectorial magnetization orientation, the input and output datasets are required. According to the Richards and Wolf diffraction theory\cite{Richards358,Youngworth2000Focusing}, the Cartesian components of the arbitrarily electric field near focus can be written as 
\begin{equation}
\mathbf{E}(\rho, \phi,z )=A\int_0^\alpha\int_0^{2\pi}\sin\theta\sqrt{\cos\theta}\mathbf{T}(\theta,\varphi)\mathbf{E}_{in}(\theta, \varphi)\times\exp[jkz\cos\theta+jk\rho\sin\theta\cos(\varphi-\phi)] d\theta d\varphi,
{\label{eq1}}
\end{equation}
with 
\begin{equation}
\mathbf{T}=
\left[
\begin{matrix}
1+(\cos\theta-1)\cos^2\varphi&(\cos\theta-1)\cos\varphi\sin\varphi&-\sin\theta\cos\varphi\\
(\cos\theta-1)\cos\varphi\sin\varphi&1+(\cos\theta-1)\sin^2\varphi&-\sin\theta\cos\varphi\\
\sin\theta\cos\varphi&\sin\theta\sin\varphi&\cos\theta
\end{matrix}
\right].
{\label{eq2}}
\end{equation}
In Eq.~(\ref{eq1}), $\mathbf{E}_{in}(\theta,\varphi)$ denotes the incident electric field and $\mathbf{T}(\theta, \varphi)$ implies the polarization transformation matrix due to the depolarization effect under the tight focusing condition. Besides, $\theta$
and $\varphi$ are the spherical coordinates in the pupil space, and $\rho$, $\phi$ and $z$ are the cylindrical coordinates in the focal region.
When the focal fields impinge on the nonabsorbing and isotropic MO film, 
the light-induced magnetization on the MO film can be described by the IFE\cite{PhysRevLett.15.190}
\begin{equation}
\mathbf {M}=j\gamma\mathbf {E}\times\mathbf{E^*}.
{\label{eq3}}
\end{equation}
In Eq.~(\ref{eq3}), $\gamma$ is magneto-optical constant.  $\mathbf{E}$ and $\mathbf{E^*}$ indicate the focal electric field and its complex conjugate, respectively. Substituting Eq.~(\ref{eq1}) into Eq.~(\ref{eq3}), the magnetization induced by tightly focusing arbitrary incident light can be  garnered.

\section{Simulated results based on machine learning}
As we all know, we can directly calculate the magnetization for arbitrary incident light. But alike some complex mathematic functions, the inverse function may be difficult to obtain or even nonexistent. To achieve the prescribed magnetization distribution, it is not easy to find the proper incident light. In order to solve this problem, the machine learning inverse design is adopted. The preacquired magnetizations and incident vector beams are regarded as input and output datasets, respectively. The input and output datasets are trained by our MANN model. In the MANN model, the general architecture is composed of an input layer with three input elements, an output layer with three output elements, and  four hidden layers with 500 hidden neurons in each layer. The input datasets are normalized (minimal to maximal scaling) before being fed to the training model. A rectified linear unit is used as the activation function. The uniform distribution to initialize all weights for neurons in each layer is adopted. For this typical regression problem, the mean square error (MSE) is chosen as the loss function. In order to get the optimal weights for the MANN model, the Gradient Descent is selected as the optimizer.

~~According to the vector diffraction theory and the IFE on the MO film, we calculate the magnetizations for 250, 000 incident beam samples. These data samples are split into two subdatasets: training (200, 000) and testing (50, 000). It is noted that the testing datasets are not used in the training, which is completely blind. In order to estimate the testing data, the hyperparameters such as the batch size, learning rate and iteration are needed to adjust during the training. When the batch size is chosen as a larger value, the required number of iterations rapidly decreases and the speed of convergence is fastly accelerated. Moreover, the convergence tendency is along the dropping direction and the corresponding oscillations decrease. When the learning rate is selected as a larger value, the best state may be missed. On the contrary, if the learning rate is enough small, the best state can be reached at the expense of a large amount of time. Therefore, the learning rate should be selected as a moderate value. The convergence result is improved with the increase of the iterations. Fortunately, these hyperparameters are easily tuned. By choosing the optimal hyperparameters, the better performance can be achieved.

\begin{figure*}[htbp]
\centering
\includegraphics[width=\linewidth]{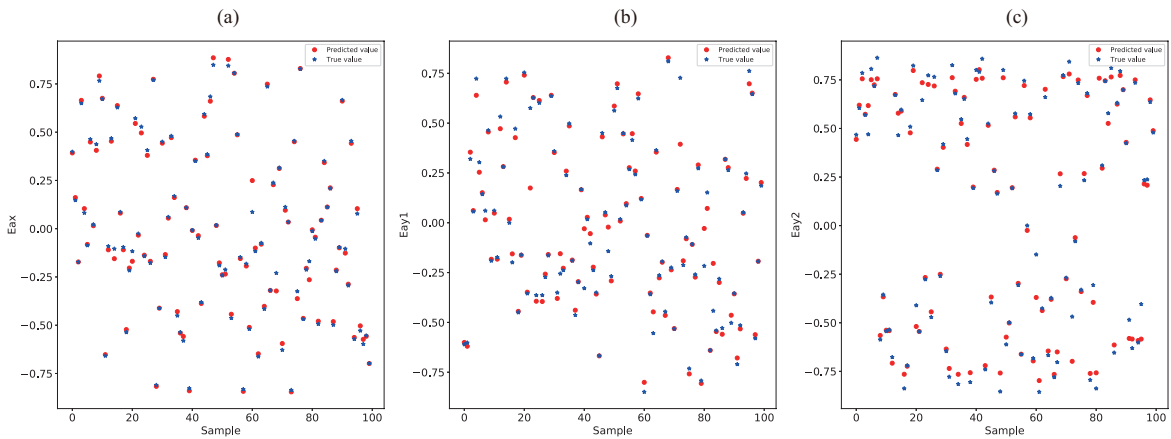}
\caption{Comparison of the predicted values and the true values in the 100 datas randomly extracting from the testing dataset. (a), (b) and (c) are for $E_{ax}$, $E_{ay1}$ and $E_{ay2}$, respectively.}
\label{fig2}
\end{figure*}

\begin{figure}[htbp]
\centering
\includegraphics[width=0.5\linewidth]{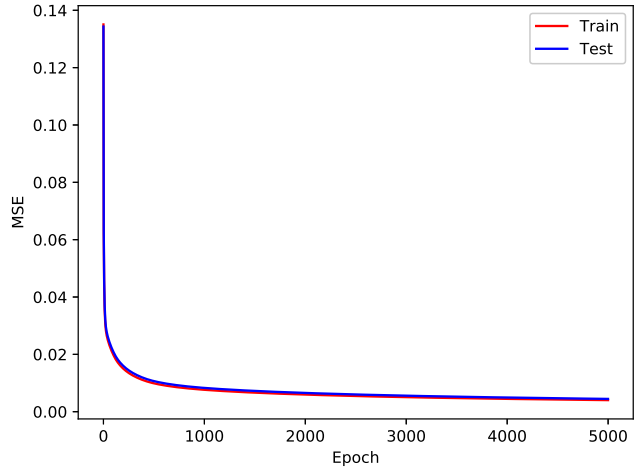}
\caption{Loss values represented by the mean squared error (MSE) in the training (red) and testing (blue) datasets as a function of the iteration indicated by epoch.}
\label{fig3}
\end{figure}

~~In our optimizing MANN model, the proper hyperparameters batch size, learning rate and the number of iteration are chosen as 1000, 0.001 and 5000, respectively. By making use of the trained machine learning model, the predicted values for the testing datasets are given. The predicted values of the amplitude parameters for the random 100 incident beam samples are presented in Fig.~\ref{fig2}. 
In Fig.~\ref{fig2}(a)-(c), both  true values and predicted values for the amplitude parameters $E_{ax}$, $E_{ay1}$ and $E_{ay2}$ are plotted. In our calculation, the absolute  values of $E_{ax}$, $E_{ay1}$ and $E_{ay2}$ are all confined to [0, $\sqrt{2}/2$].  From Fig.~\ref{fig2}, it is easily found that predicted values are close to the  true values. The training and testing loss values represented by the MSE are plotted in Fig.~\ref{fig3}, when the number of iterations increases. It is seen from Fig.~\ref{fig3} that the loss values on the training and testing datasets show the same variation trend as the iteration increases. Firstly, the  loss values decline rapidly when the iterations are the range of [0, 100]. Subsequently, the loss values decrease in a quite low speed. At last, loss values do not decrease and converge in a stable small value (about $0.2\%$).
%The training and testing loss values represented by the MSE are plotted in Fig.~\ref{fig3}, when the number of iteration increases. It is seen from Fig.~\ref{fig3} that the loss values on the training and testing datasets show the same variation trends as the iteration increases. Firstly, the  loss values decline rapidly when the iterations are the range of [0, 100]. Subsequently, the loss values decrease in a quite low speed. At last, loss values do not decrease and converge in a stable small value (about $0.2\%$).
Combining with the results in the Fig.~\ref{fig2} and Fig.~\ref{fig3}, the reasonability and availability of our MANN model are sufficiently proved.

\begin{figure}
\centering
\includegraphics[width=0.5\linewidth]{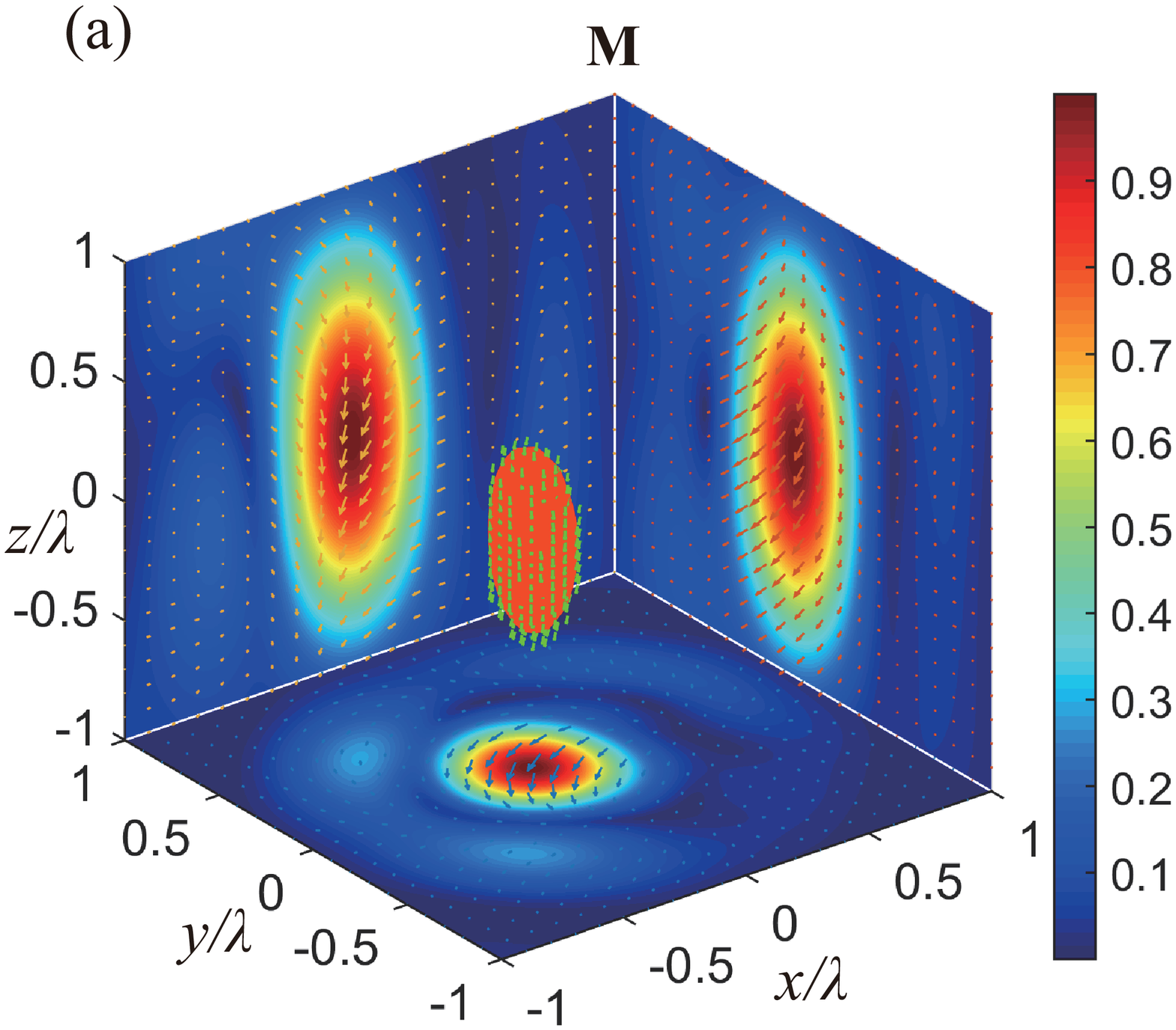}  \\
\includegraphics[width=0.5\linewidth]{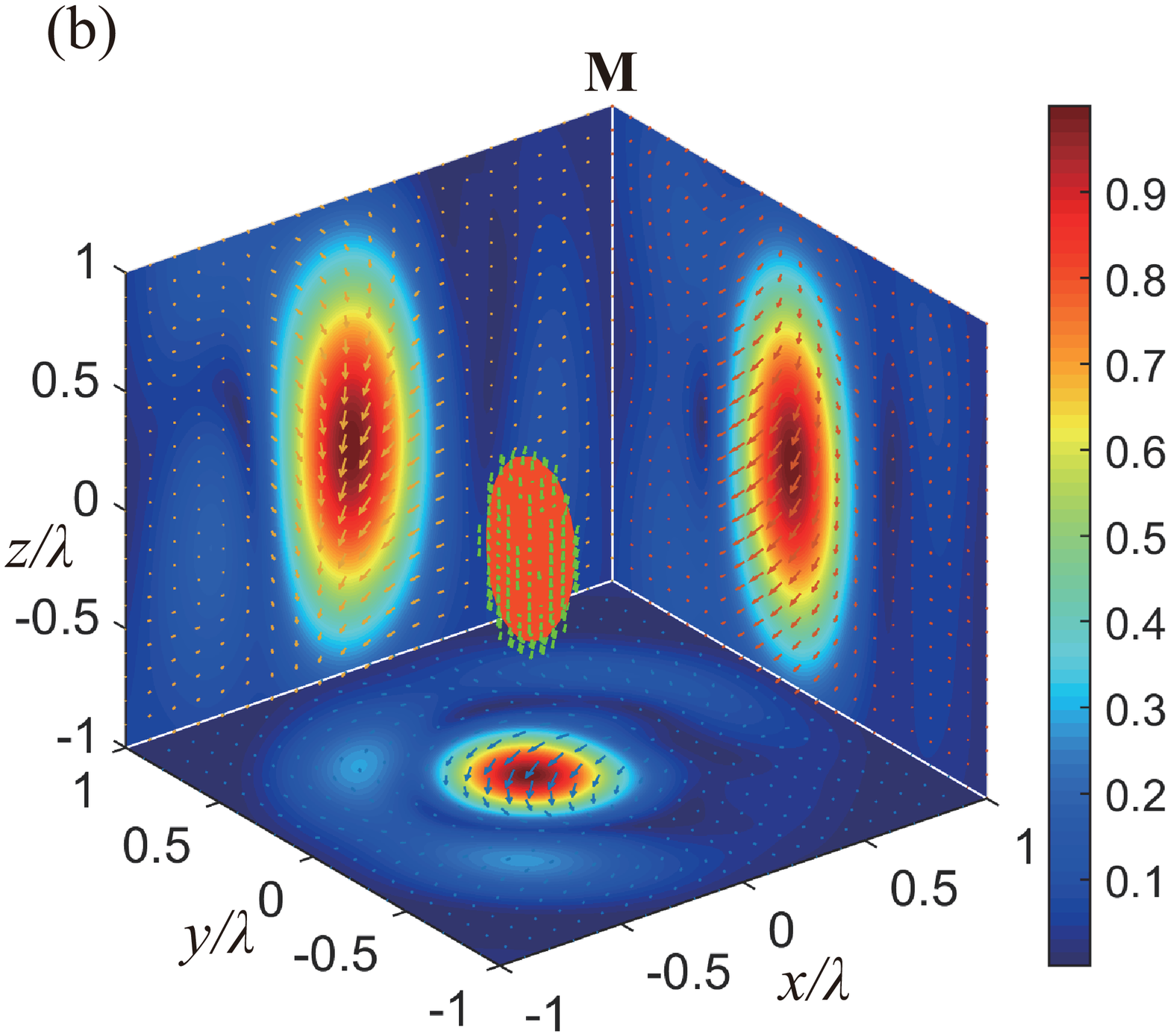}
\caption{(a)-(b) The projections of magnetization intensity (color) and magnetization orientation (arrows) along orthogonal planes as well as the 3D isosurface (color) with the 80 percentages of the maximum magnetization intensity and the corresponding surrounding 3D magnetization orientations (arrows) for the
testing values in one sample and its corresponding estimated values.}
\label{fig4}
\end{figure}

~~To give a direct and vivid demonstration of the power in our MANN model, we give an example. Here, we choose one sample datas. The electric amplitude factors are selected as $E_{ax}=0.398$, $E_{ay1}=-0.611$ and $E_{ay2}=0.467$. Meanwhile, the estimated electric amplitude factors are $E_{ax}=0.385$, $E_{ay1}=-0.601$ and $E_{ay2}=0.464$. These two groups of values are nearly close.
According to these two groups of electric amplitude factors, the projections of magnetization intensities and magnetization orientations along orthogonal planes as well as the 3D isosurface with the 80 percentages of the maximum magnetization intensity and the corresponding surrounding 3D magnetization orientations are depicted in Fig.~\ref{fig4}. It is easily found from Fig.~\ref{fig4} that the orientations of the magnetization are distributed in three orthogonal planes. And as seen from the 3D isosurface magnetization intensity surrounded by magnetization orientation, the magnetization orientations near the focus are along 3D direction.
%According to these two groups of electric amplitude factors, the projections of magnetization intensities and magnetization orientations along orthogonal planes as well as the 3D isosurface with the 80 percentages of the maximum magnetization intensity and the corresponding surrounding 3D magnetization orientations are depicted in Fig.~\ref{fig4}. 
%It is easily found from Fig.~\ref{fig4} that the orientations of the magnetization are distributed in three orthogonal planes. And as seen from the 3D isosurface magnetization intensity surrounded by magnetization orientation, the total magnetization vector near the focus is along 3D direction.
More importantly, the magnetization distributions of intensities and orientations in Fig.~\ref{fig4}(a)-(b) are almost the same except for some trivial side lobes.

~~The 3D magnetization orientations in Fig.~\ref{fig4}(a) and Fig.~\ref{fig4}(b) are vectors. Mathematically, the similarities and discrepancies between two vectors can be estimated by $S_m=\cos(\widehat{\mathbf{a}, \mathbf{b}})=\frac{\mathbf{a}\cdot\mathbf{b}}{\left|\mathbf{a}\right| \left|\mathbf{b}\right|}$. Here, $\widehat{\mathbf{a}, \mathbf{b}}$ denotes the angle between two vectors ($\mathbf{a}$ and $\mathbf{b}$). When the angle between the two vectors is equal to 0, the cosine of the angle is 1. In this case, the directions of the two vectors are identical. When the angle between the two vectors is equal to 90 degrees, the cosine of the angle is 0. In other words, the two vectors are perpendicular. Therefore, $S_m$ ranging from 0 to 1 can be applied to estimate the similarities between two different vectors. The true magnetization orientation ($M_x$, $M_y$, $M_z$) in the focus is (-0.559, -0.479, -0.560). And the estimated magnetization orientation is ($M_x$, $M_y$, $M_z$)=(-0.570, -0.475, -0.546). The cosine of the angle between the true and estimated magnetization orientations can be calculated by our definition $S_m$. And the calculated $S_m$ is equal to 0.9998. This value is in close proximity to 1, which means that the true and estimated magnetization orientations are nearly identical.
%The 3D magnetization orientations in Fig4.(a) and Fig4.(b) are vectors. Mathematically, the similarities and discrepancies between two vectors can be estimated by $S_m=\cos(\widehat{\mathbf{a}, \mathbf{b}})=\frac{\mathbf{a}\cdot\mathbf{b}}{\left|\mathbf{a}\right| \left|\mathbf{b}\right|}$. Here, $\widehat{\mathbf{a}, \mathbf{b}}$ denotes the angle between two vectors ($\mathbf{a}$ and $\mathbf{b}$). When the angle between the two vectors is equal to 0, the cosine of the angle is 1. In this case, the directions of the two vectors are identical. When the angle between the two vectors is equal to 90 degrees, the cosine of the angle is 0. In other words, the two vectors are perpendicular. Therefore, $S_m$ ranging from 0 to 1 can be used to estimate the similarities between two different vectors. The true magnetization orientation ($M_x$, $M_y$, $M_z$) in the focus is (-0.559, -0.479, -0.560). And the estimated magnetization orientation is ($M_x$, $M_y$, $M_z$)=(-0.570, -0.475, -0.546). The cosine of the angle between the true and estimated magnetization orientations can be calculated by our definition $S_m$. And the calculated $S_m$ is equal to 0.9998. This value is very close to 1, which means that the true and estimated magnetization orientations are nearly identical. 

~~To further perform a reliable examination of the quality about the trained neural network, some different input configurations which are not contained in the training and testing datasets are used as input values. The corresponding electric fields are predicted by our well trained neural network. According to the IFE in the MO film, the final estimated magnetization orientations are calculated. The desired magnetization orientaions ($\mathbf{M}$, $M_x$, $M_y$ and $M_z$) and the estimated magnetization orientations ($M_{xe}$, $M_{ye}$, $M_{ze}$) are shown in table 1. Besides, the similarities ($S_m$) between the desired magnetization orientaions and the estimated magnetization orientations are also calculated. As seen from this table, most values of $S_m$ are reached to above 0.96, which are very chose to 1. Only one $S_m$ is equal to about 0.85. It is common phenomenon to appear a small amount of worse samples because that the machine learning guarantee the overall estimated effect for all training samples not only for a small number of samples. As a whole, the estimated magnetization orientations are almost indentical to the orientations of the target magnatization.
%~~To further perform a reliable examination of the quality about the training neural network, some different input configurations which are not contained in the training and testing datasets are used as input values. The corresponding electric fields are predicted by our trianing neural network. According to the IFE in the MO film, the final estimated magnetization orientations are calculated. The desired magnetization orientaions ($\mathbf{M}$, $M_x$, $M_y$ and $M_z$) and the estimated magnetization orientations ($M_{xe}$, $M_{ye}$, $M_{ze}$) orientations are shown in table 1. Besides, the similarities ($S_m$) between the desired magnetization orientaions and the estimated magnetization orientations are also calculated. As seen from this table, most of the $S_m$ is reached to above 0.96, which is very chose to 1. Only one $S_m$ is equal to about 0.85, which is less than 0.90. It is common phenomenon to appear some worse samples. This is because that the machine learning ganrantee the overall estimated effect for all training samples not for some small samples. As a whole, the estimated magnetization orientations are almost indentical to the orientation of the target magnatization.
%{\color{red}
\begin{table}[htbp]
\centering
\caption{\bf Comparision between the target magnetization orientations ($\mathbf{M}$, $M_x$, $M_y$ and $M_z$)  and the estimated magnetization orientations ($M_{xe}$, $M_{ye}$, $M_{ze}$). Here, $S_m$ denotes the similarity between them.}
\begin{tabular}{|c|c|c|c|c|c|}
\hline
Targets $\mathbf{M}(\alpha,\beta)$ & $\mathbf{M}(\pi,\pi/2)$ & $\mathbf{M}(\pi/3,\pi/6)$ & $\mathbf{M}(4\pi/3,\pi/6)$ & $\mathbf{M}(5\pi/6,2\pi/3)$ & $\mathbf{M}(7\pi/6,\pi/6)$\\
\hline
$M_x$ & -1 & $\frac{1}{4}$ & $-\frac{1}{4}$ & $-\frac{3}{4}$ & $-\frac{\sqrt{3}}{4}$ \\
\hline
$M_y$ & 0 & $\frac{\sqrt{3}}{4}$  & $\frac{\sqrt{3}}{4}$ & $\frac{\sqrt{3}}{4}$ & $-\frac{1}{4}$ \\
\hline
$M_z$ & 0 & $\frac{\sqrt{3}}{2}$ & $\frac{\sqrt{3}}{2}$ & $-\frac{1}{2}$ & $\frac{\sqrt{3}}{2}$ \\
\hline
$M_{xp}$  & -1.0000 & 0.2500 & -0.2500 & -0.7500 & -0.4330 \\
\hline
$M_{yp}$  & 0 & 0.4330 & 0.4330 & 0.4330 & -0.2500 \\
\hline
$M_{zp}$  & 0 & 0.8660 & 0.8660 & -0.5000 & 0.8660 \\
\hline
$S_m$  & 1.0000 & 0.9679 & 0.9626 & 0.9996 & 0.8514 \\
\hline
\end{tabular}
\end{table}
This result again proves that the predictions based on our MANN model are quite accurate.

~~By adopting our proposed method, the required incident beam can be predicted for an expected magnetization orientation. In the experiment, the complex incident beam can be produced by vectorial optical field generator\cite{Han:13}. Through the modulation of the incident beam, the expected magnetization orientation can be realized under the tight focusing condition. Magnetization spot arrays with controllable 3D orientations are produced by point-to-point scanning. Each spot in these magnetization distributions has independent magnetization orientation and can be controlled arbitrarily. The controlable magnetization orientation based on our machine learning algorithm greatly facilite the formation of magnetic holography, which strengthens ultrahigh information security. The magnetic holography can be used as identity verification for a credit card with magnetic stripe. On account of the freedom of storage of the 3D magnetization orientation, the realized multistate could advance the development in multiplexing magnetization storage and information processing technologies, and it may also find applications in the detection of magnetic particles.

\section{Conclusion}
In conclusion, we put up with the machine learning inverse design to achieve vectorial magnetization orientation. According to the RMs under tightly focusing condition and the IFE on the nonabsorbing MO film, a structured incident beam composed of four kinds of beams is focused to four polarized focal fields, allowing to produce a magnetization spot with 3D orientation. Based on the vector diffraction theory and the IFE, two groups of datasets comprised of the magnetizations and incident beams are garnered. These two groups of datasets are fed into the MANN model and trained by optimizing hyperparameters. After training, it is revealed that the predicted values are almost close to the testing values for the majority of samples. The loss value decreases with the increase of the iteration and finally converges to the stable value ($0.2\%$). Importantly, the calculated magnetization distributions based on the testing and estimated values of one incident beam sample are almost the same. Besides, some prospects on how to experimentally realize the 3D magnetization and its interesting applications are discussed. It is highly expected that the proposed machine learning inverse design can open up broad potentials in magnetization shapings containing structures and orientations. 
% Experimental section

\medskip
%\textbf{Supporting Information} \par %Please delete the Suppporting Information statement if it is not applicable. Please supply Supporting Information in another file. Supporting information should not be provided in .tex format
%Supporting Information is available from the Wiley Online Library or from the author.

% Acknowledgements
\medskip
\textbf{Acknowledgements} \par %delete if not applicable))
This work was supported by the National Natural Science Foundation of China (12004155, 11974258, 11904152, 11604236, 61575139, 61865009, 61927813, 60504052); Key Research and Development (R$\&$D) Projects of Shanxi Province, China (201903D121127); Scientific and Technological Innovation Programs of Higher Education Institutions in Shanxi (2019L0151).

\textbf{Conﬂict of Interest} \par %delete if not applicable))
The authors declare no conflict of interest.

% References
\medskip

% Use the following code if you wish to generate your bibliography with BibTeX;
% replace the string "MSP-template" below with the name(s) of
% the BibTeX data base(s) you want to use.
% The resulting bibliography-output (the content of the .bbl file)
% must be pasted back into this file before submission.
% Please also include your BibTeX data base file(s) in your submission
% so that we can re-run BibTeX if necessary.
%
\bibliographystyle{}

\end{document}